\documentclass[preprint,12pt]{elsarticle}

\usepackage{amssymb}
\usepackage{amsmath}
\usepackage{graphicx}
\usepackage{multirow}
\usepackage{hyperref}
\usepackage{cleveref}
\usepackage[utf8]{inputenc} 
\usepackage[T1]{fontenc}    
\usepackage{hyperref}       
\usepackage{url}            
\usepackage{booktabs}       
\usepackage{amsfonts}       
\usepackage{nicefrac}       
\usepackage{microtype}      
\usepackage{xcolor}         
\usepackage{caption}

\journal{Medical Image Analysis}

\begin{document}

\begin{frontmatter}



\title{Multi-Contrast MRI Motion Correction via Parameter-Informed Disentanglement and Adaptive Experts} 


\author[1]{Honglin Xiong} 
\author[1]{Yuxian Tang}
\author[1]{Feng Li}
\author[1]{Yulin Wang}
\author[2]{Lei Xiang}
\author[1,2,3]{Dinggang Shen}
\author[1,3]{Qian Wang}
\ead{qianwang@shanghaitech.edu.cn}

\affiliation[1]{organization={School of Biomedical Engineering \& State Key Laboratory of Advanced Medical Materials and Devices, ShanghaiTech University},
            city={Shanghai},
            postcode={201210}, 
            state={Shanghai},
            country={China}}
\affiliation[2]{organization={Shanghai United Imaging Intelligence Co. Ltd.},
            city={Shanghai},
            postcode={200230}, 
            state={Shanghai},
            country={China}}
\affiliation[3]{organization={Shanghai Clinical Research and Trial Center},
            city={Shanghai},
            postcode={201210}, 
            state={Shanghai},
            country={China}}
\begin{abstract}
Motion artifacts in magnetic resonance imaging (MRI) degrade diagnostic reliability.  
Existing deep learning methods are typically contrast-specific and fail to generalize across diverse modalities and artifact severities. 
We propose a unified framework combining parameter-informed contrast disentanglement with severity-aware adaptive correction. 
ScanCLIP, pretrained on over 30,000 MRI text–image pairs, derives contrast embeddings from acquisition parameters to disentangle contrast style from anatomical content, yielding contrast-free features.  
A Vision Transformer then estimates motion severity and routes features through a Mixture-of-Experts network, enabling targeted artifact correction.  
A dual-pathway decoder reconstructs both the clean image and residual artifact map, enforcing image-space consistency.  
On IXI and HCP benchmarks, our method improves PSNR by 0.75 dB and SSIM by up to 2.79\% over state-of-the-art approaches, with larger gains at higher artifact severities.  
It further demonstrates robust zero-shot generalization on real-world clinical data acquired with unseen scanning parameters, where existing methods either fail to remove artifacts or introduce additional distortions.  
\end{abstract}

\begin{graphicalabstract}
\includegraphics[width=\textwidth]{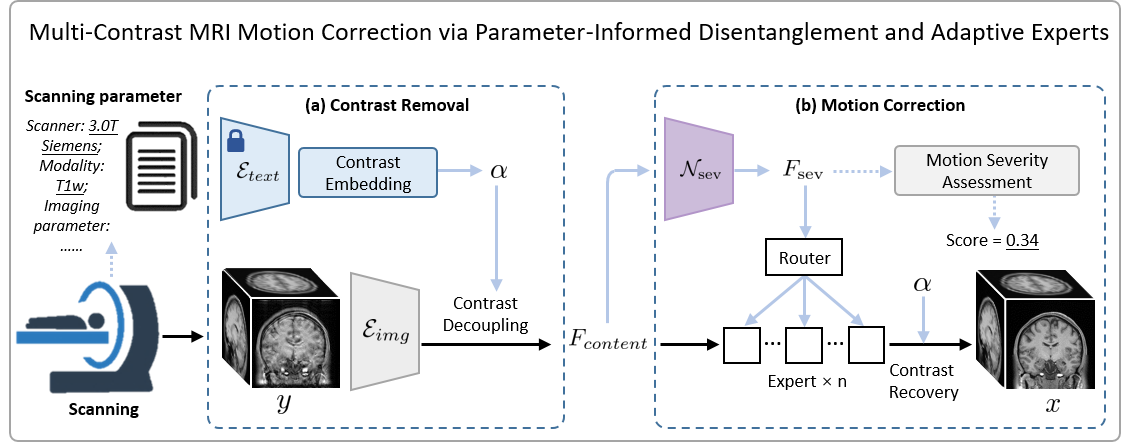}

\end{graphicalabstract}

\begin{highlights}
\item A unified framework for robust motion correction across diverse MRI contrasts and severities.
\item A severity-aware Mixture-of-Experts architecture enables adaptive, targeted artifact correction.
\item Demonstrates superior performance and robust zero-shot generalization on real-world clinical data where other SOTA methods fail.
\end{highlights}


\end{frontmatter}



\section{Introduction}
\label{sec:introduction}

Magnetic Resonance Imaging (MRI) is a cornerstone of modern medical diagnostics, prized for its exceptional soft tissue contrast and non-invasive visualization capabilities that are critical for early disease detection and treatment planning. However, the diagnostic utility of MRI is frequently compromised by motion artifacts stemming from patient movement during acquisition. These artifacts can severely degrade image quality, obscure vital anatomical details, potentially lead to misdiagnosis, and necessitate costly, inconvenient repeat scans~\citep{zaitsev2015motion}.

Numerous strategies have been developed to mitigate MRI motion artifacts~\citep{Zaitsev2015review}. Among these, retrospective motion correction (MoCo) techniques are particularly attractive due to their non-invasive nature and computational efficiency, allowing them to be applied post-acquisition without altering scanning protocols~\citep{Loktyushin2013blind}. In recent years, deep learning has emerged as a powerful paradigm for MoCo. Data-driven methods, including Convolutional Neural Networks (CNNs) and Generative Adversarial Networks (GANs), have shown promise in learning mappings from motion-corrupted to clean images~\citep{Johnson2019cGan,Kustner2019DL}. More sophisticated architectures, such as those incorporating self-attention mechanisms~\citep{tsai2023motion} to capture long-range dependencies, and diffusion models~\citep{oh2023annealed,xie2022measurement}, have further advanced the state-of-the-art.

Despite these advancements, the clinical translation of current deep learning MoCo models faces significant hurdles. A primary challenge is their limited generalization across diverse MRI contrasts (e.g., T1w, T2w, PDw). Models optimized for one contrast often perform poorly on others because image appearance and artifact manifestation vary substantially across sequences. For instance, T2w sequences, with their typically longer echo times (TE), are inherently more susceptible to motion than T1w sequences. This lack of robustness restricts their practical utility, especially as clinical acquisition routinely involves multiple MRI sequences.
Although contemporary ``all-in-one'' restoration models~\citep{potlapalli2023PromptIR} have shown promise in handling multiple types of inputs, they are largely developed for natural image degradation. In contrast, MRI is governed by complex acquisition physics, and motion artifacts are closely tied to sequence-specific parameters. 
In addition, motion artifacts in MRI span a wide range of severity. Since mild and severe motion corruption differ substantially in both appearance and restoration difficulty, effective correction requires severity-aware adaptation rather than a single fixed mapping. These challenges motivate the development of an MRI-specific framework that is both contrast-robust and severity-adaptive.

To develop a robustly translatable tool, a MoCo framework must overcome two intertwined challenges: (1) the vast variability in artifact appearance across different MRI contrasts, and (2) the wide spectrum of artifact severity. Addressing both is crucial, as it would enable a single model to be reliably deployed across diverse clinical scanning protocols, eliminating the need for multiple specialized models.

However, a central obstacle lies in the fact that conventional models often learn features where artifact patterns are deeply entangled with contrast-specific anatomical information, limiting their generalization. To break this entanglement, our strategy is to disentangle contrast from anatomical content, building upon our prior work on contrast synthesis~\citep{xiong2025learning}.  Specifically, we propose a unified framework that leverages MRI scan parameters to computationally remove contrast influence, thereby isolating a contrast-free representation of the underlying anatomy with its motion corruption. These contrast-free features then serve as the common input for subsequent stages of motion severity assessment and motion correction.

To realize these objectives, our unified framework proceeds in three synergistic stages. 
First, the contrast disentanglement stage leverages our ScanCLIP model, which pretrained on over 30{,}000 MRI image–parameter pairs, to derive sequence-specific contrast embeddings from textual acquisition parameters. 
By factoring out these embeddings from the input features, we obtain contrast-free content representations that standardize anatomical and motion-related information across diverse MRI sequences. 
Second, these standardized features are processed by a Vision Transformer (ViT) to estimate motion severity, and the resulting severity-aware representation is used to guide the adaptive Mixture-of-Experts (MoE) architecture. 
Through this severity-informed routing, the MoE selectively activates specialized sub-experts to disentangle clean anatomical structures from motion-induced errors. 
Finally, a dual-pathway decoding strategy re-entangles the original contrast embeddings to simultaneously reconstruct the clean image and the residual artifact map, enforcing image-space consistency for high-fidelity restoration.

\section{Related Work}

\paragraph{Image Restoration}
Image restoration, a long-standing challenge in computer vision and image processing, seeks to recover clean images from degraded observations. Early model-based approaches, such as total variation regularization~\citep{rudin1992nonlinear} and dictionary learning for sparse representations~\citep{elad2006image}, laid the foundation for this field. The advent of deep learning, particularly CNN-based methods like DnCNN~\citep{zhang2017beyond}, revolutionized restoration by learning complex mappings from corrupted to clean images. More recent architectures have further advanced performance: Transformer-based designs such as the Swin Transformer~\citep{liu2021swin} capture global context effectively, while lightweight networks like NAFNet~\citep{chen2022simple} achieve state-of-the-art efficiency. Diffusion models~\citep{ho2020denoising,rombach2022high} have introduced a powerful paradigm for high-fidelity synthesis and artifact removal. In parallel, explorations into foundation models such as CLIP~\citep{radford2021learning} have inspired text-guided or prompt-based restoration strategies, enabling more context-aware outcomes~\citep{potlapalli2023PromptIR}.  

Despite these advances, most approaches are designed for natural images and overlook the unique challenges of medical imaging. MRI, in particular, involves complex acquisition physics, where artifacts are tightly coupled with sequence-specific parameters and patient physiology. This domain-specific complexity necessitates tailored solutions that go beyond generic restoration pipelines.

\paragraph{Motion Correction in MRI}
Motion artifacts remain a major impediment to MRI diagnostic quality. Mitigation strategies have evolved from prospective approaches, such as real-time motion tracking during acquisition~\citep{maclaren2013prospective}, to retrospective techniques that correct images post-acquisition~\citep{zaitsev2015motion}. Traditional retrospective methods, often based on image registration~\citep{jenkinson2002improved}, can be effective for mild motion but struggle with severe or non-rigid artifacts, particularly in multi-contrast scenarios.  
Deep learning has emerged as the dominant paradigm for retrospective MRI motion correction~\citep{duffy2021deep,chen2020self,yaman2020self}, offering data-driven mappings that outperform conventional approaches. Recent innovations increasingly emphasize robustness across multiple contrasts and the integration of auxiliary information~\citep{wang2022transformer}. Importantly, metadata and acquisition parameters have been explored to enhance correction accuracy and adaptability across diverse scanning protocols~\citep{lin2021deep}. However, effectively leveraging these parameters to achieve consistent correction across wide-ranging motion severities and contrasts remains an open challenge that our work directly addresses.

\section{Method}
\label{sec:method}

\begin{figure}[htbp]
    \centering
    \includegraphics[width=\linewidth]{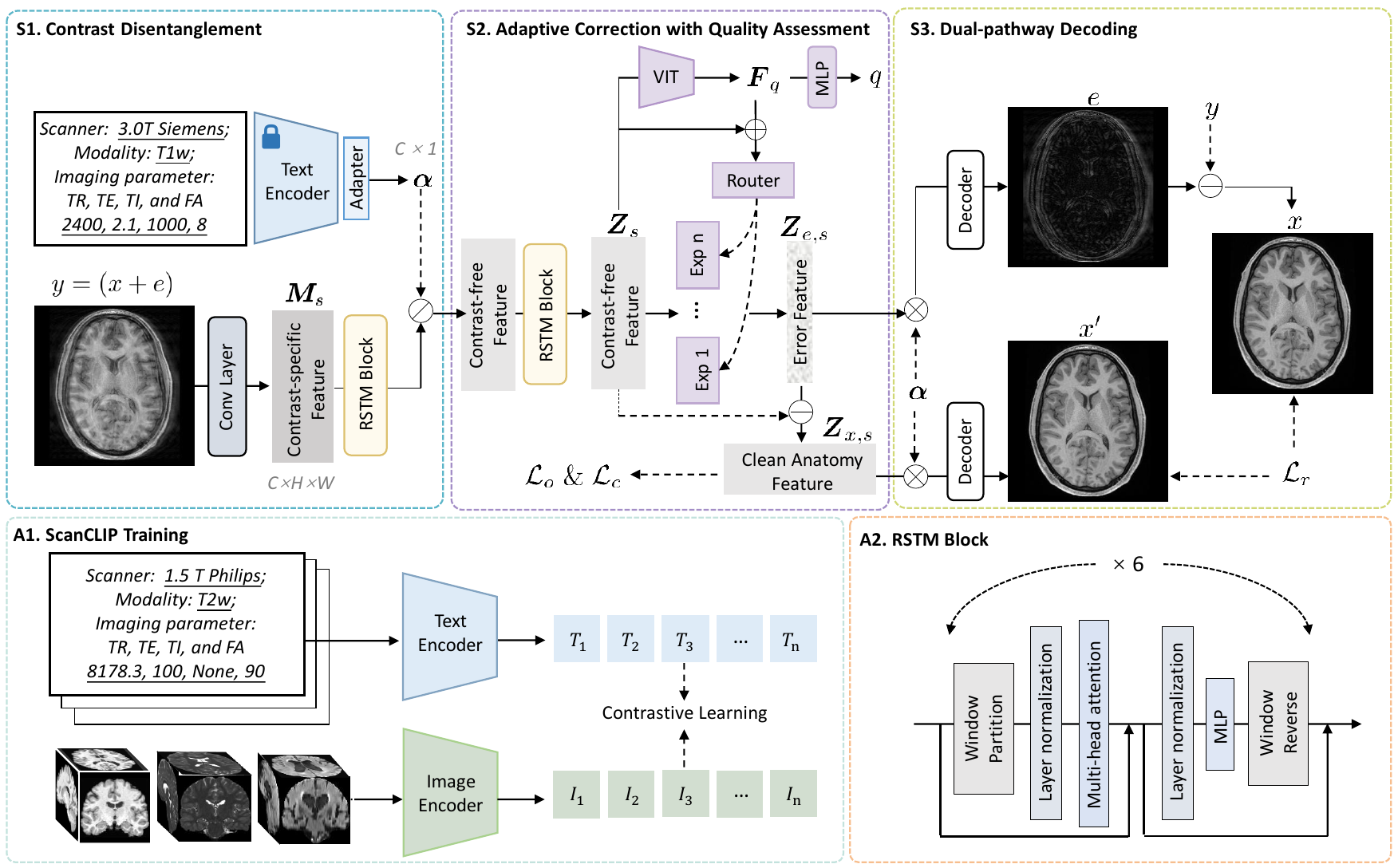}
    \caption{Overview of our unified framework for multi-contrast MRI motion correction. 
The pipeline consists of three stages: (S1) text-informed contrast disentanglement; 
(S2) adaptive correction with motion severity assessment; 
and (S3) dual-pathway decoding that reconstructs the clean image and the residual artifact map.
}

    \label{fig:method}
\end{figure}

\subsection{{Problem Definition and Overall Framework}}
\label{sec:problem_definition}

\paragraph{{Problem Definition}}
Let $x \in \mathbb{R}^{H \times W}$ denote a clean, artifact-free MRI slice, where $H$ and $W$ represent the image height and width. In clinical acquisitions, patient motion introduces phase errors in the \textit{k}-space, which manifest as structured, non-linear corruptions (e.g., ghosting and blurring) in the reconstructed magnitude images. A widely adopted strategy in deep learning is to formulate restoration as a residual learning problem, where the corrupted image $y$ is modeled as
\begin{equation}
y = x + e,
\end{equation}
with $e$ representing the error map. In this formulation, the network is tasked with estimating $e$, ensuring that the anatomical details of $x$ are preserved during correction.

Multi-contrast MRI protocols produce images of the same anatomy with varying appearances, governed by acquisition parameters such as TR and Echo Time (TE). These parameters dictate global intensity distributions and tissue contrast. We hypothesize that, in a high-dimensional feature space, the representation {$\boldsymbol{M}$} of an acquired image $y$ can be disentangled into two components: a sequence-dependent contrast embedding $\boldsymbol{\alpha}$ and a contrast-free content feature map $\boldsymbol{Z}$ that captures spatial structures. 
Following our prior work~\cite{xiong2025learning}, we model this relationship as
\begin{equation}
\boldsymbol{M}(y) = \boldsymbol{\alpha} \odot \boldsymbol{Z}(y),
\end{equation}
where $\boldsymbol{\alpha}$ is derived exclusively from textual scan parameters via the pretrained ScanCLIP. Because $\boldsymbol{\alpha}$ is dependent on acquisition parameters, $\boldsymbol{\alpha}$ encodes only the intended global contrast ``style'' while remaining unaffected by stochastic patient motion. 

This disentanglement provides a robust foundation: by factoring out $\boldsymbol{\alpha}$, we eliminate sequence-specific intensity variations while preserving the structural topology and spatial patterns in $\boldsymbol{Z}$ that characterize both the underlying anatomy $x$ and the motion-induced error $e$. Building on this normalization, our framework is designed to achieve two key objectives:  
\begin{enumerate}
    \item To computationally remove the influence of MRI contrast $\boldsymbol{\alpha}$, thereby isolating contrast-free content features $\boldsymbol{Z}$ that standardize diverse contrasts while retaining explicit motion artifacts.  
    \item To feed these standardized features into an adaptive network that further disentangles them into clean anatomy features $\boldsymbol{Z}_x$ and error features $\boldsymbol{Z}_e$, enabling targeted correction and faithful reconstruction of the clean image.  
\end{enumerate}

\paragraph{{Overall Framework}}

To realize these objectives, we propose a unified framework that integrates contrast disentanglement with adaptive MoCo. 
As illustrated in Figure~\ref{fig:method}, the framework proceeds in three stages. 
In the first stage (S1), the contrast-coupled image feature $\boldsymbol{M}_s$ (with $s$ denoting the input sequence) is normalized using the sequence-dependent contrast embedding $\boldsymbol{\alpha}$ derived from ScanCLIP, yielding a contrast-free content representation $\boldsymbol{Z}_s$ (Section~\ref{sec:contrast_disentanglement}). 
In the second stage (S2), a ViT processes $\boldsymbol{Z}_s$ to estimate the motion severity and produces a severity-aware feature vector $\boldsymbol{F}_q$. 
This severity representation dynamically routes $\boldsymbol{Z}_s$ to appropriate sub-experts within the MoE architecture, enabling the disentanglement of clean anatomical structures $\boldsymbol{Z}_{x,s}$ and motion-induced errors $\boldsymbol{Z}_{e,s}$ (Section~\ref{sec:adaptive_correction_with_quality_assessment}). 
Finally (S3), a dual-pathway decoding strategy reintroduces the original contrast embedding $\boldsymbol{\alpha}$ into $\boldsymbol{Z}_{x,s}$ and $\boldsymbol{Z}_{e,s}$ to reconstruct the clean image ${x}'$ and the residual artifact map ${e}$, respectively, while enforcing image-space consistency between ${x}'$ and $x = {y} - {e}$ for high-fidelity restoration (Section~\ref{sec:dual_pathway_decoding}).

\subsection{Contrast Disentanglement with ScanCLIP}
\label{sec:contrast_disentanglement}

{
As shown in Figure~\ref{fig:method}-S1, the first stage of our framework leverages contrast embeddings $\boldsymbol{\alpha}$ generated by the text encoder from the pretrained ScanCLIP model to remove sequence-specific intensity variations from $\boldsymbol{M}_s$, thereby yielding contrast-free content features from $\boldsymbol{Z}_s$. 
Specifically, given an input MRI image $y$ from a specific sequence, a convolutional image encoder (shared across all sequences) produces a $C$-channel feature map $\boldsymbol{M}_s = (M_s^1, \ldots, M_s^C) \in \mathbb{R}^{H \times W \times C}$.
At the same time, the ScanCLIP text encoder processes the acquisition parameters to yield a contrast embedding $\boldsymbol{\alpha} = (\alpha^1, \ldots, \alpha^C)$.
Each element $\alpha^k$ quantifies the influence of the sequence-specific contrast on the $k$-th channel of $\boldsymbol{M}_s$. To decouple contrast information, we normalize each channel by its corresponding embedding coefficient:
\begin{equation}
Z_s^k = \frac{M_s^k}{\alpha^k + \epsilon}, \quad k = 1, \ldots, C,
\label{eq:contrast_normalization}
\end{equation}
where $\epsilon$ is a small constant for numerical stability~\citep{xiong2025learning}. 
The resulting contrast-free features $\boldsymbol{Z}_s = (Z_s^1, \ldots, Z_s^C)$  serve as the standardized input for the subsequent motion correction.
}

The training process of ScanCLIP is depicted in Figure~\ref{fig:method}-A1. Following our prior work~\citep{wang2025toward}, ScanCLIP was trained on over 30,000 MRI image-parameter pairs, learning robust associations between textual acquisition parameters and corresponding image contrasts. Specifically, adopting a standard CLIP training paradigm, the model utilizes a Transformer-based text encoder to process the textual parameters and a CNN-based image encoder to process the MRI scans. A contrastive learning objective is then optimized to align their respective feature representations within a shared latent space. In this work, we employ only the frozen text encoder from ScanCLIP to derive a sequence-specific contrast embedding ${\boldsymbol{\alpha}}$ for each input sequence. Because ${\boldsymbol{\alpha}}$ is determined entirely by fixed acquisition parameters (e.g., TR, TE, TI, FA), it encapsulates the intended global intensity style of the MRI contrast while remaining unaffected by stochastic patient motion. This property makes ${\boldsymbol{\alpha}}$ an effective guide for disentangling contrast style from anatomical content and motion artifacts.

\subsection{Severity-Aware Adaptive Correction}
\label{sec:adaptive_correction_with_quality_assessment}

\paragraph{{Quality Assessment}}
{In Figure~\ref{fig:method}-S2,} after obtaining the contrast-free features $\boldsymbol{Z}_s$, the next step is to estimate the severity of motion degradation so that subsequent correction can be adaptively guided. We employ a ViT that operates on non-overlapping $32 \times 32$ patches of $\boldsymbol{Z}_s$. The model consists of six transformer layers with an embedding dimension of 512 and eight attention heads per layer.

The ViT produces two outputs: a scalar quality score $q \in [0,1]$ and a severity feature vector $\boldsymbol{F}_q$. The score $q$ reflects the degree of motion corruption and serves as the supervisory target for training the severity estimator. To construct this score, we combine three complementary full-reference image-quality metrics computed between the corrupted image $y$ and its clean reference $x$: Peak Signal-to-Noise Ratio (PSNR), Structural Similarity Index (SSIM)~\citep{wang2004image}, and Visual Information Fidelity (VIF)~\citep{sheikh2006image}. PSNR captures global fidelity via intensity differences, SSIM measures luminance, contrast, and structural consistency, and VIF quantifies the shared information between $y$ and $x$ in the wavelet domain.

\begin{figure}[htbp]
    \centering
    \includegraphics[width=0.9\linewidth]{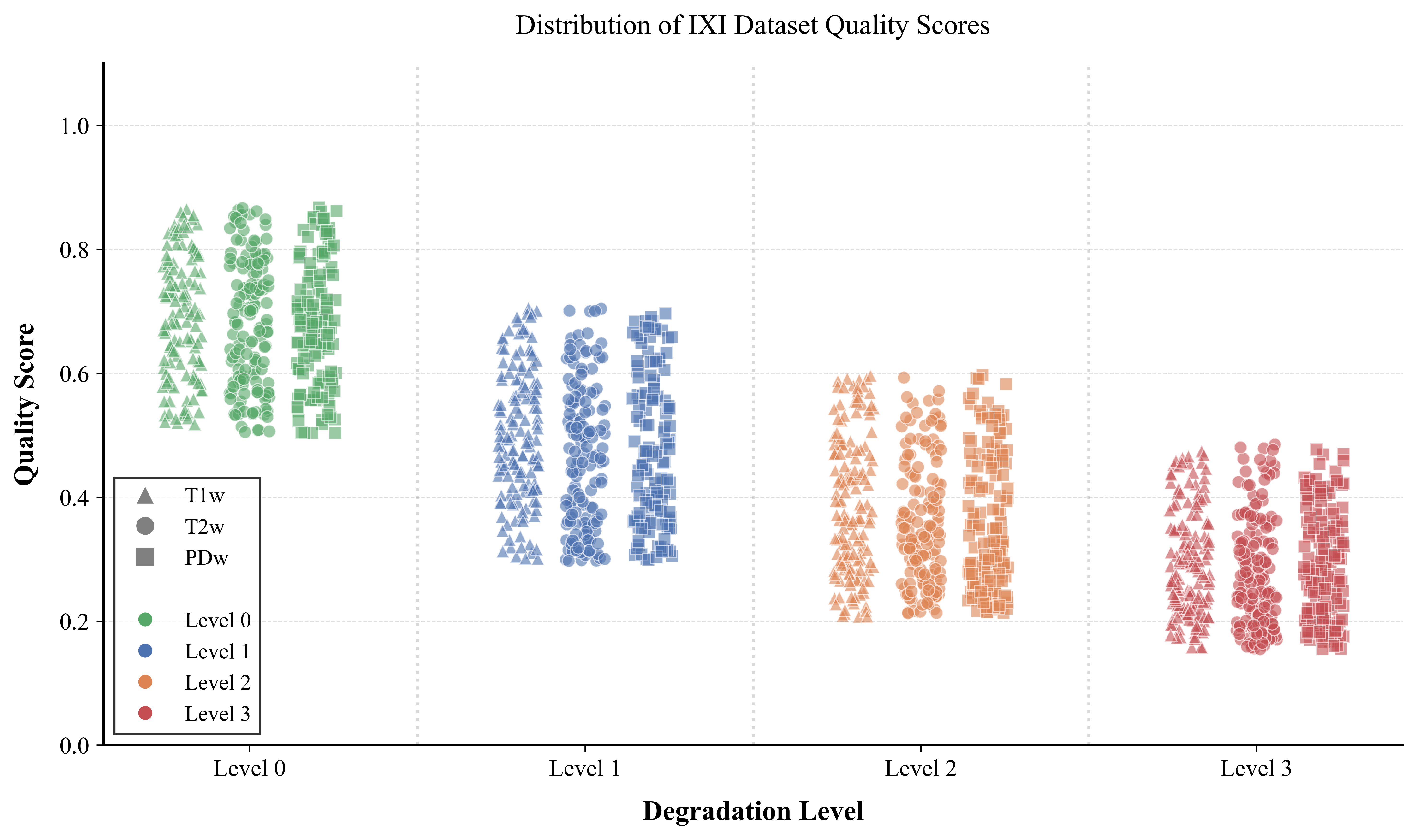}
    \caption{Distribution of quality scores ($q$) on the IXI dataset. The scatter plot illustrates the distribution of the calculated composite quality scores ($q$) against different motion severity levels.}
    \label{fig:ixi_scatter}
\end{figure}

Because these metrics operate on different numerical scales, each is normalized to $[0,1]$ using a clipped function
\begin{equation}
\mathcal{N}(v; v_{\min}, v_{\max}) =
\begin{cases}
0, & v \le v_{\min}, \\
\frac{v - v_{\min}}{v_{\max} - v_{\min}}, & v_{\min} < v < v_{\max}, \\
1, & v \ge v_{\max},
\end{cases}
\end{equation}
with empirically determined bounds based on the distribution of our training data: $[15,34]$ for PSNR, $[0.4,1.0]$ for SSIM, and $[0,1]$ for VIF. After clipping, the three metrics are combined into a single quality indicator by computing a weighted average of the normalized PSNR, SSIM, and VIF values, assigning weights of $0.3$, $0.4$, and $0.3$, respectively, to reflect their relative perceptual importance. This aggregated value is then passed through a gamma correction with $\gamma = 1.5$, which amplifies differences in the low-quality regime and improves sensitivity to severe motion artifacts. The final output of this process is the composite score $q$, a smooth and monotonic measure of motion severity that remains consistent across different MRI contrasts. Its empirical distribution on the IXI dataset is shown in Figure~\ref{fig:ixi_scatter}, validating its ability to separate different levels of motion degradation and we observe similar trends extend robustly to additional MRI datasets.

The ViT is trained to regress this composite score $q$ from $\boldsymbol{Z}_s$. In parallel, the internal CLS token representation before the regression head is extracted as the severity feature vector $\boldsymbol{F}_q$. This vector captures high-level semantic cues about the type and extent of motion artifacts and is subsequently used to guide the adaptive MoE correction network. 
Together, $q$ and $\boldsymbol{F}_q$ enable the framework to reason about motion severity and dynamically adjust the correction strategy.

\paragraph{{Adaptive Correction}}

The motion correction operates on the contrast-free features $\boldsymbol{Z}_s$ and adapts its behavior according to the estimated motion severity. To achieve this, our network architecture relies on the Residual Swin Transformer Module (RSTM) as its foundational building block. These RSTM blocks extract deep hierarchical features from the input. As illustrated in Figure \ref{fig:method}-A2, to enable adaptive correction, each RSTM block within the severity-aware expert network replaces its standard feed-forward network (FFN) sub-layer with an MoE layer. This modification allows the network to dynamically select specialized processing pathways tailored to different degradation levels.

As the contrast-free features $\boldsymbol{Z}_s$ progress through the RSTM layers, the resulting intermediate feature representations enter the MoE FFN layer. To incorporate motion awareness, these intermediate features are concatenated with the severity feature vector $\boldsymbol{F}_q$ 
(obtained from the motion severity assessment).
This concatenation forms a unified guidance representation that serves as the shared input to both the routing network and the specialized experts.

The MoE layer consists of a main expert and a set of sub-experts (10 in our implementation). 
A lightweight router network processes the unified guidance representation and produces a normalized set of routing weights, which determine the contribution of each sub-expert. Each expert receives the same guidance representation as input. Their individual outputs are then combined according to the router weights and passed through a GeLU activation. The final output of the MoE FFN layer is obtained by fusing this activated combination of sub-expert outputs with the output generated by the main expert. This design allows the network to emphasize different expert pathways depending on the motion severity explicitly encoded in $\boldsymbol{F}_q$.

The MoE-enhanced RSTM blocks are trained to predict the residual error features $\boldsymbol{Z}_e$ from the input contrast-free features $\boldsymbol{Z}_s$. The estimated clean anatomical features, $\boldsymbol{Z}_x$, are then explicitly obtained by subtracting the predicted error features from the input:
\begin{equation}
\boldsymbol{Z}_{x,s} = \boldsymbol{Z}_s - \boldsymbol{Z}_{e,s}.
\end{equation}
This residual decomposition yields separate representations for the underlying clean anatomy and the motion-induced artifacts. 

To further refine the representation of the clean anatomical structures, two constraints are imposed on $\boldsymbol{Z}_{x,s}$ during training. First, a consistency loss $\mathcal{L}_{c}$ enforces similarity between clean features estimated from different sequence contrasts of the same subject, promoting contrast invariance. Specifically, given clean anatomical features $\boldsymbol{Z}_{x,s_1}$ and $\boldsymbol{Z}_{x,s_2}$ extracted from two different contrasts (e.g., T1w and T2w) of the same slice, the consistency loss is formulated as the $L_1$ distance between them:
\begin{equation}
\label{eq:consistency_loss}
\mathcal{L}_{c} = \mathbb{E}_{s_1 \neq s_2} \left[ \left\| \boldsymbol{Z}_{x,s_1} - \boldsymbol{Z}_{x,s_2} \right\|_1 \right],
\end{equation}
where the expectation is taken over all available valid contrast pairs within the training batch. This formulation ensures that the extracted structural information remains invariant regardless of the input modality.
Second, an orthogonalization loss $\mathcal{L}_o$ \citep{xiong2025learning} encourages the $C$ channels of $\boldsymbol{Z}_{x,s}$ to form diverse, factorized bases:
\begin{equation}
\mathcal{L}_o = \frac{\sum_{k=1}^{C} \sum_{k' \neq k}^{C} \left| ({Z}_{x,s}^k)^T {Z}_{x,s}^{k'} \right|}{\sum_{j=1}^{C} \|{Z}_{x,s}^j\|_2^2},
\label{eq:orthogonalization}
\end{equation}
where ${Z}_{x,s}^k$ is the $k$-th channel of $\boldsymbol{Z}_{x,s}$. Together, this subtraction-based decomposition and the regularizing constraints yield robust, disentangled features that form the basis for the subsequent contrast re-modulation and image-space reconstruction stage.

\subsection{Dual-pathway Decoding}
\label{sec:dual_pathway_decoding}

To achieve a more robust correction that maximally preserves anatomical detail, we employ a novel dual-pathway decoding strategy (Figure~\ref{fig:method}-S3). 
After disentanglement, the clean and artifact features are re-entangled with the contrast embedding $\boldsymbol{\alpha}$, which is the inverse operation to Eq.~\eqref{eq:contrast_normalization} following \citep{xiong2025learning}. 
They are subsequently processed by a shared decoder $\mathcal{D}$ composed of $3 \times 3$ convolutional layers. 
The decoder operates through two parallel pathways. 
First, in the direct clean estimation pathway, $\mathcal{D}$ transforms the re-entangled clean anatomical features to produce an estimate of the clean image, denoted as $x'$. 
Second, in the residual error estimation pathway, the same decoder processes the re-entangled artifact features to estimate an error map $e$, from which the final corrected image is obtained by residual subtraction:
\begin{equation}
x = y - e.
\label{eq:final_corrected_image}
\end{equation}
To ensure spatial and predictive consistency between these two pathways, we introduce an image-space reconstruction loss, $\mathcal{L}_{r}$, which minimizes the $L_1$ difference between the directly decoded clean image $x'$ and the residual-corrected output $x$:
\begin{equation}
\mathcal{L}_{r} = \| x' - x \|_1 = \| x' - (y - e) \|_1.
\label{eq:recon_cons_loss}
\end{equation}

Beyond enforcing consistency loss in image space, this dual-pathway design also plays a critical role in aligning the latent-space decomposition with the objectives of the second stage (Figure~\ref{fig:method}-S2) in our method. 
That is, by explicitly decoding both $\boldsymbol{Z}_x$ and $\boldsymbol{Z}_e$, the network is encouraged to maintain a clean separation between anatomical and artifact features, ensuring that the latent representations used for severity-aware routing remain semantically meaningful and structurally coherent. 
This alignment substantially strengthens the disentanglement learned in the expert network and contributes to the overall robustness and accuracy of the motion correction process as revealed by our experiments.

\section{Experiments}
\label{sec:exp}
\subsection{Experimental Setup}
\subsubsection{Datasets}
\label{sec:datasets}
\paragraph{ScanCLIP Pretraining Dataset}
Our ScanCLIP model was pretrained on over 30,000 brain MRI volumes. These data were sourced from diverse global institutions and public datasets, including OASIS~\citep{lamontagne2019oasis}, HCP~\citep{van2013wu}, BCP~\citep{howell2019unc}, ADNI~\citep{petersen2010alzheimer}, AIBL~\citep{ellis2009australian}, BraTS2021~\citep{baid2021rsna}, and our internal datasets. These extensive and diverse datasets allow ScanCLIP to learn robust association between textual scan parameters and corresponding image characteristics.

\paragraph{Motion Correction Datasets}

Our motion correction framework was trained and evaluated using a combination of public datasets and an in-house cohort. 
The dataset configurations used in our experiments are summarized below. 
\begin{itemize}
    \item \textbf{IXI Dataset:} We selected 200 subjects across T1w, T2w, and PDw contrasts, yielding 21{,}600 training slices, 2{,}400 validation slices, and 4{,}800 testing slices.
    \item \textbf{HCP Dataset:} We used 200 subjects with T1w and T2w scans, providing 16{,}000 training slices, 1{,}600 validation slices, and 3{,}200 testing slices. 
    
\item \textbf{In-house Real-Motion Dataset:} 
To evaluate real-world cross-contrast capability of our  framework, we curated an in-house dataset consisting of eight subjects with visible, authentic motion artifacts. 
Crucially, these scans were acquired on a United Imaging uMR~890 scanner. The scanning parameters (TR = 3000~ms, TE = 452.0~ms, TI = 1100~ms, flip angle = $57^{\circ}$,  isotropic voxel size of $0.8$~mm, matrix of $240 \times 280$, and 240 slices) were close to T2w, yet fully unseen by any (pre-)training data used in this work.  
This setup provides a stringent test of our model's zero-shot generalization to unseen contrast styles. 
\end{itemize}

For the IXI and HCP datasets, we generated motion corruptions by retrospectively simulating head motion in $k$-space following the protocol of \citep{duffy2021retrospective}. 
This procedure models patient movement by applying a sequence of 3D rigid-body transformations to different segments of the $k$-space acquisition, thereby producing realistic ghosting and blurring patterns that reflect motion occurring during the scan. 
Specifically, we simulated three representative motion trajectories: (1) \emph{piecewise constant} motion, in which the head remains in a fixed pose for a block of phase-encoding lines before abruptly transitioning to a new pose; (2) \emph{piecewise transient} motion, which introduces brief, rapid perturbations between otherwise stable poses; and (3) \emph{smooth Gaussian} motion, where motion parameters evolve continuously according to a Gaussian process. 
Following the severity definitions in \citep{duffy2021retrospective}, four levels of corruption (0--3) were created by varying the proportion of motion-affected phase-encoding lines: 0\%--10\% for Level~0, 10\%--20\% for Level~1, 20\%--30\% for Level~2, and 30\%--40\% for Level~3. 
To preserve fundamental contrast and structural integrity, the central 7\% of $k$-space was kept motion-free across all severity levels.

\subsubsection{Training Procedures}
\label{sec::train}

\paragraph{ScanCLIP Pretraining}
ScanCLIP was pretrained for 100 epochs using a warm-up--cosine learning rate schedule, where the learning rate increased to $5\times10^{-5}$ over the first 20 epochs and then decayed to zero. 
A batch size of 27 was used to ensure sufficient negative samples across parameter classes, and optimization employed Adam with $\beta_1=0.5$ and $\beta_2=0.999$. 
The final checkpoint was used to initialize the text encoder for all downstream experiments. 
The total pretraining time was approximately 40 hours on a single NVIDIA A100 GPU.

\paragraph{Motion Correction Model Training}
The motion correction network was trained for 80 epochs on a single NVIDIA A100 GPU, requiring roughly 45 hours. 
Training used the Adam optimizer with an initial learning rate of $1\times10^{-4}$, which was decayed by a factor of 0.9 every 20 epochs via a multi-step scheduler.

\subsection{Results on the Public Datasets}
\label{sec:moco}
To comprehensively evaluate our framework, we conducted extensive comparisons with state-of-the-art methods. These include Transformer-based image restoration models SwinIR~\citep{liang2021SwinIR} and Restormer~\citep{zamir2022restormer}, the diffusion-model-based approach ResShift~\citep{yue2023ResShift}, the all-in-one medical image restoration model AMIR~\citep{yang2024all}, and the prompt-based PromptIR~\citep{potlapalli2023PromptIR}. Experiments were performed on multi-contrast data from the HCP and IXI datasets. 
\begin{table}[htbp]
    \centering
    \caption{Quantitative evaluation of motion correction performance on IXI and HCP datasets. The ``Ours'', highlighted in bold, shows statistically significant improvements over the other compared methods (\textit{p} < 0.005 in paired t-tests) across the various conditions and metrics presented.}
    \label{tab:combined_results}
    \resizebox{\textwidth}{!}{%
    \begin{tabular}{ccccccccccccccc} 
    \toprule
    \multirow{2}{*}{Dataset} & \multirow{2}{*}{Contrast} & \multirow{2}{*}{\begin{tabular}[c]{@{}c@{}}Motion\\ Level\end{tabular}} & \multicolumn{6}{c}{PSNR (dB) $\uparrow$} & \multicolumn{6}{c}{SSIM (\%) $\uparrow$} \\
    \cmidrule(lr){4-9} \cmidrule(lr){10-15}
     & & & SwinIR & Restormer & ResShift & PromptIR & AMIR & Ours & SwinIR & Restormer & ResShift & PromptIR & AMIR & Ours \\ 
    \midrule
    \multirow{12}{*}{IXI} & \multirow{4}{*}{T1w}& 0 & 42.70 & 42.53 & 42.21 & 43.37 & 44.14 & \textbf{44.92} & 99.20 & 99.08 & 98.93 & 99.39 & 99.48 & \textbf{99.75} \\
     &  & 1 & 38.67 & 38.56 & 38.14 & 39.27 & 39.44 & \textbf{40.43} & 98.54 & 98.22 & 98.04 & 98.74 & 98.85 & \textbf{99.40} \\
     &  & 2 & 36.24 & 35.88 & 35.61 & 36.84 & 37.09 & \textbf{38.36} & 97.63 & 97.41 & 97.30 & 97.83 & 98.00 & \textbf{98.86} \\
     &  & 3 & 34.09 & 33.81 & 33.58 & 34.83 & 34.69 & \textbf{35.65} & 96.91 & 96.77 & 96.43 & 97.11 & 97.30 & \textbf{98.39} \\ 
    \cmidrule(lr){2-15} 
     & \multirow{4}{*}{T2w}& 0 & 42.20 & 41.91 & 41.76 & 43.17 & 42.83 & \textbf{43.60} & 98.80 & 98.63 & 98.37 & 99.01 & 99.13 & \textbf{99.58} \\
     &  & 1 & 37.81 & 37.69 & 37.38 & 38.41 & 38.54 & \textbf{39.05} & 97.75 & 97.44 & 97.26 & 97.95 & 98.12 & \textbf{99.00} \\
     &  & 2 & 34.63 & 34.25 & 34.08 & 35.23 & 35.36 & \textbf{35.94} & 96.37 & 96.26 & 95.89 & 96.57 & 96.75 & \textbf{98.17} \\
     &  & 3 & 32.43 & 32.26 & 31.91 & 33.03 & 33.31 & \textbf{33.85} & 95.10 & 94.88 & 94.75 & 95.30 & 95.49 & \textbf{97.37} \\ 
    \cmidrule(lr){2-15} 
     & \multirow{4}{*}{PDw} 
        & 0 & 42.30 & 42.06 & 41.67 & 43.92 & 43.60 & \textbf{44.90} & 99.08 & 98.92 & 98.61 & 99.18 & 99.28 & \textbf{99.63} \\
     &  & 1 & 36.90 & 36.57 & 36.42 & 38.71 & 38.81 & \textbf{39.47} & 98.01 & 97.89 & 97.64 & 98.11 & 98.27 & \textbf{99.03} \\
     &  & 2 & 34.19 & 34.01 & 33.68 & 36.20 & 36.38 & \textbf{37.09} & 97.03 & 96.65 & 96.52 & 97.13 & 97.29 & \textbf{98.50} \\
     &  & 3 & 31.69 & 31.32 & 31.14 & 33.82 & 33.86 & \textbf{34.55} & 95.61 & 95.42 & 95.11 & 95.71 & 95.90 & \textbf{97.65} \\ 
    \midrule
    \midrule
    \multirow{8}{*}{HCP} & \multirow{4}{*}{T1w}& 0 & 37.31 & 37.16 & 36.79 & 38.08 & 38.06 & \textbf{38.70} & 96.56 & 96.22 & 96.06 & 96.86 & 97.15 & \textbf{98.89} \\
     &  & 1 & 33.12 & 32.83 & 32.61 & 34.44 & 34.48 & \textbf{35.30} & 95.01 & 94.88 & 94.59 & 95.31 & 95.54 & \textbf{98.07} \\
     &  & 2 & 29.56 & 29.45 & 29.11 & 31.09 & 31.31 & \textbf{32.22} & 92.71 & 92.48 & 92.35 & 93.01 & 93.29 & \textbf{96.85} \\
     &  & 3 & 28.19 & 27.81 & 27.65 & 29.61 & 29.87 & \textbf{30.64} & 91.27 & 91.13 & 90.75 & 91.57 & 91.84 & \textbf{95.93} \\ 
    \cmidrule(lr){2-15} 
     & \multirow{4}{*}{T2w}& 0 & 38.22 & 37.95 & 37.62 & 38.93 & 39.11 & \textbf{40.01} & 96.57 & 96.41 & 96.21 & 96.87 & 97.27 & \textbf{98.94} \\
     &  & 1 & 34.42 & 34.29 & 33.88 & 35.57 & 35.54 & \textbf{36.24} & 95.27 & 94.94 & 94.82 & 95.57 & 96.10 & \textbf{98.20} \\
     &  & 2 & 31.60 & 31.25 & 31.06 & 32.78 & 32.83 & \textbf{33.49} & 93.71 & 93.59 & 93.31 & 94.01 & 94.28 & \textbf{97.22} \\
     &  & 3 & 29.74 & 29.53 & 29.16 & 30.77 & 30.88 & \textbf{31.56} & 92.03 & 91.76 & 91.60 & 92.33 & 92.52 & \textbf{96.17} \\ 
    \bottomrule
    \end{tabular}%
    }
\end{table}

The quantitative results, detailed in Table~\ref{tab:combined_results}, demonstrate the superior performance of our proposed method. Across both the IXI and HCP datasets, our approach consistently achieved the highest PSNR and SSIM scores for all evaluated MRI contrasts and motion severity levels. 
Specifically, on the IXI dataset, our method showed an average improvement of approximately $0.75$ dB in PSNR and $0.90\%$ in SSIM over the next best performing method across all tested conditions. For the HCP dataset, the average improvements were approximately $0.75$ dB in PSNR and $2.79\%$ in SSIM.

\begin{figure}[htbp]
  \includegraphics[width=\textwidth]{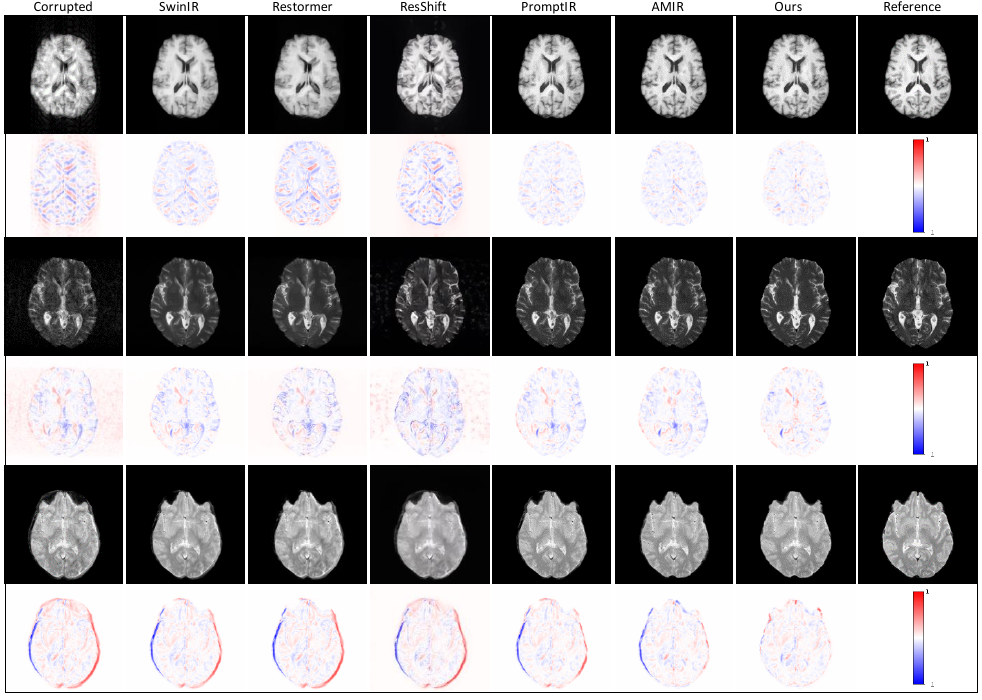} 
  \caption{Qualitative motion correction results on the IXI dataset. Our method is compared against other methods on three contrasts (T1w, T2w and PDw) with severe (level 3) motion artifacts. }
  \label{fig:ixi}
\end{figure}

\begin{figure}[htbp]
    \centering
    \includegraphics[width=\textwidth]{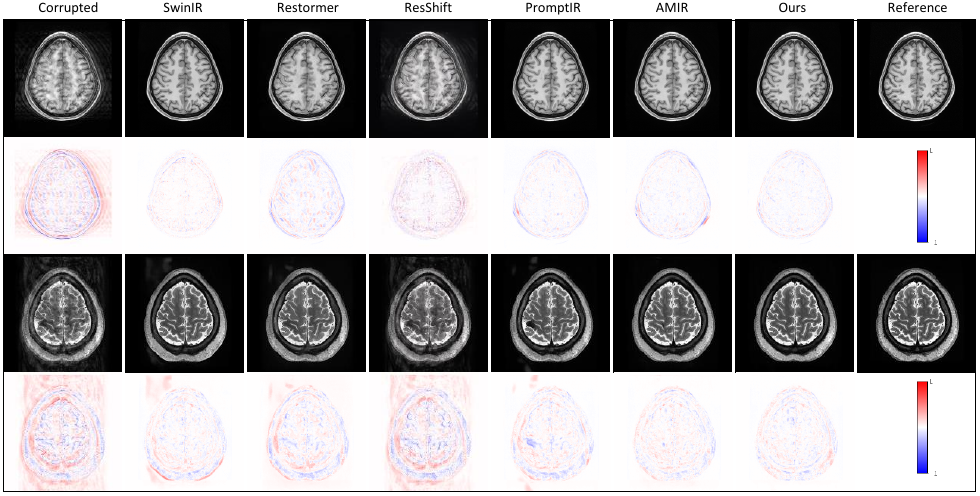}
    \caption{Qualitative motion correction results on the HCP dataset. Our method is compared against other methods on two contrasts (T1w and T2w) with severe (level 3) motion artifacts. }
    \label{fig:hcp_moco} 
\end{figure}

Qualitative results further corroborate these quantitative findings. Figure~\ref{fig:ixi} visualizes correction performance on the IXI dataset, showcasing results for T1w, T2w, and PDw images under severe (Level 3) motion artifacts. As illustrated, our method yields images with substantially reduced artifacts and better preservation of anatomical details compared to other methods. Similarly, Figure~\ref{fig:hcp_moco} presents correction results for the HCP dataset, further demonstrating our method's robust performance across different data sources. In summary, our approach consistently produces images with clearer anatomical delineation and fewer residual artifacts than the compared methods, particularly in challenging high-artifact scenarios.

\subsection{Performance on the Real-World Dataset}
\label{sec:real_data_performance}
\begin{figure}[htbp]
    \includegraphics[width=\textwidth]{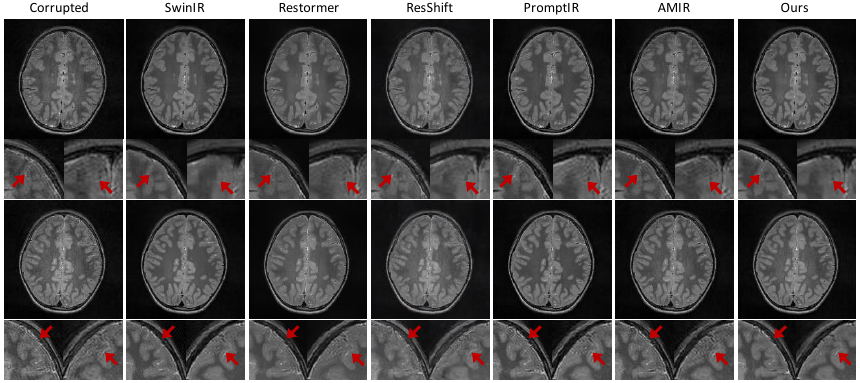}
    \caption{Qualitative comparison of motion correction on challenging real-world MRI data. Our method effectively reduces artifacts and preserves anatomical details on full-slice examples.}
    \label{fig:ccbd}
\end{figure}

\begin{figure}[htbp]
    \centering
    \includegraphics[width=\textwidth]{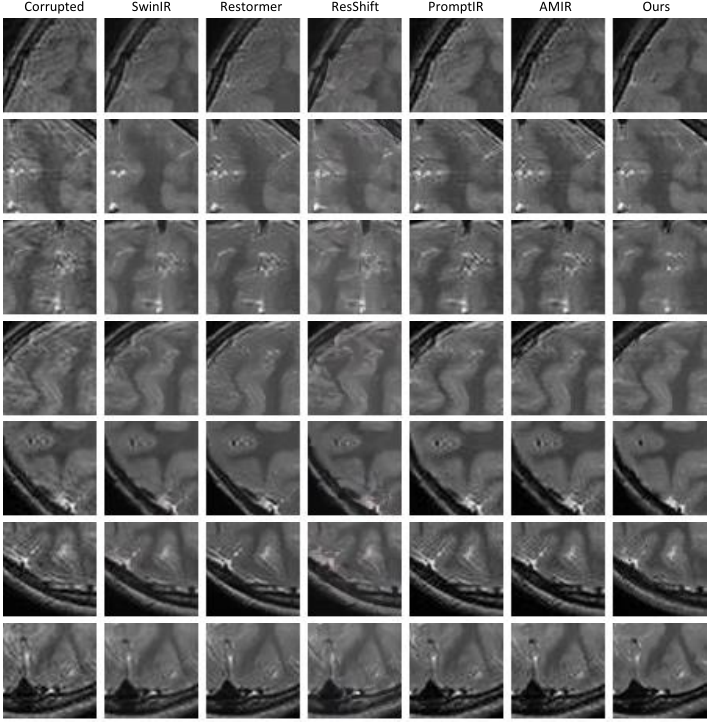} 
    \caption{Further examples of motion correction on the in-house real clinical dataset, supplementing Figure~\ref{fig:ccbd}. To facilitate detailed comparison, only zoomed-in regions are shown, illustrating our method's consistent effectiveness in reducing complex, real-world motion artifacts and enhancing anatomical clarity.} 
    \label{fig:ccbd_more} 
\end{figure}

To assess generalization and clinical applicability, our framework was validated on an in-house dataset. This challenging dataset contains scans with real motion artifacts. Notably, its scanning parameters were {entirely unseen by ScanCLIP during pretraining}, stringently testing our model's adaptability.
Without ground-truth motion-free references, Figure~\ref{fig:ccbd} presents representative visual comparisons on full slices against state-of-the-art methods. Further examples are provided in Figure~\ref{fig:ccbd_more}, which showcases zoomed-in regions from different subjects to better highlight the fine details of artifact removal and anatomical preservation.

Visual inspection of these results reveals the limitations of several contemporary methods. For instance, prominent techniques like Restormer, PromptIR, and AMIR often failed to effectively eliminate pervasive motion artifacts, leaving significant residual distortions. More critically, other approaches, including SwinIR and ResShift, were observed to introduce new anatomical inaccuracies (e.g., distorted white matter structures) during attempted correction. 
In contrast, our proposed method demonstrated superior artifact suppression capabilities while faithfully preserving underlying anatomical details. As seen in both the full slices in Figure~\ref{fig:ccbd} and the detailed regions in Figure~\ref{fig:ccbd_more}, our framework effectively addressed the motion corruption without introducing such adverse structural alterations. These findings on real-world clinical data, particularly with scanning parameters unseen during training, underscore the robust generalization performance of our unified framework. The ability to correct genuine motion artifacts while maintaining anatomical integrity highlights its significant potential for practical clinical utility.

\subsection{Ablation Study}
\label{sec:ablation_study}

We conducted ablation experiments on the IXI dataset to quantify the contribution of each major component in our framework (Table~\ref{tab:ablation_studies_side}).

\paragraph{S1: ScanCLIP-guided contrast disentanglement}
To evaluate this component, we bypassed the first stage of ScanCLIP entirely, feeding the raw, unnormalized image features directly into the subsequent severity assessment and correction networks. Removing the contrast disentanglement stage led to a substantial PSNR drop from $38.36$~dB to $36.45$~dB. This confirms that normalizing contrast variations is essential for enabling the downstream network to focus on motion-related structures rather than sequence-dependent intensity differences.

\paragraph{S2: Adaptive Mixture-of-Experts}
Eliminating all sub-experts but leaving only the main expert (i.e., $N_e = 0$) degraded PSNR to $36.93$~dB, demonstrating the importance of severity-aware specialization. 
We further evaluated different numbers of sub-experts ($N_e \in \{5, 10, 15\}$). 
Increasing $N_e$ from 5 to 10 improved PSNR from $37.95$~dB to $38.36$~dB and SSIM from $98.65\%$ to $98.86\%$. 
A larger expert pool ($N_e = 15$) yielded only marginal gains (PSNR $38.38$~dB; SSIM essentially unchanged), indicating diminishing returns. 
Thus, $N_e = 10$ is adopted in our implementation as it provides an effective balance between representational capacity and computational efficiency.

\paragraph{S3: Dual-pathway Decoding}
To evaluate this component, we compared our dual-pathway strategy against a common single-pathway baseline. In this single-pathway setup, the decoder utilizes only one branch to directly predict the final corrected image, entirely omitting the parallel estimation of the residual error map. Replacing the dual-pathway decoder with this single-pathway alternative reduced the PSNR to $37.12$~dB. The explicit separation of clean and artifact features in our proposed design therefore provides a clearer supervisory signal, improving reconstruction fidelity.

\begin{table}[htbp]
    \centering
    \caption{Ablation study of key components on the IXI dataset (Mixed motion level). $N_e$: Number of sub-experts in MoE.}
    \label{tab:ablation_studies_side}
    \begin{tabular}{@{}lcc@{}}
        \toprule
        Config & PSNR (dB) $\uparrow$ & SSIM (\%) $\uparrow$ \\
        \midrule
        Full Model (Ours) & $38.36 \pm 0.12$ & $98.86 \pm 0.18$ \\
        \midrule
        w/o Contrast Disentanglement     & $36.45 \pm 0.25$ & $97.75 \pm 0.15$ \\
        \midrule
        $N_e=0$    & $36.93 \pm 0.22$ & $97.98 \pm 0.14$ \\
        $N_e=5$    & $37.95 \pm 0.15$ & $98.65 \pm 0.10$ \\
        $N_e=15$   & $38.38 \pm 0.11$ & $98.86 \pm 0.07$ \\
        \midrule
         w/o Dual-pathway Decoding    & $37.12 \pm 0.20$ & $98.10 \pm 0.12$ \\
        \bottomrule
    \end{tabular}
\end{table}

\subsection{Segmentation Evaluation}
\label{sec:segmentation_evaluation}
To assess the downstream clinical utility of our motion correction framework, we evaluated its impact on automated brain tissue segmentation using T1-weighted scans from the HCP dataset—a task highly sensitive to motion-induced blurring and intensity distortions.
We used FMRIB's Automated Segmentation Tool (FAST)~\citep{zhang2001segmentation} to segment cerebrospinal fluid (CSF), grey matter (GM), and white matter (WM) from images corrected by our method and competing approaches. 
Segmentations obtained from the original artifact-free scans served as the reference, and accuracy was quantified using the Dice coefficient.

\begin{figure}[htbp]
    \centering
    \includegraphics[width=\textwidth]{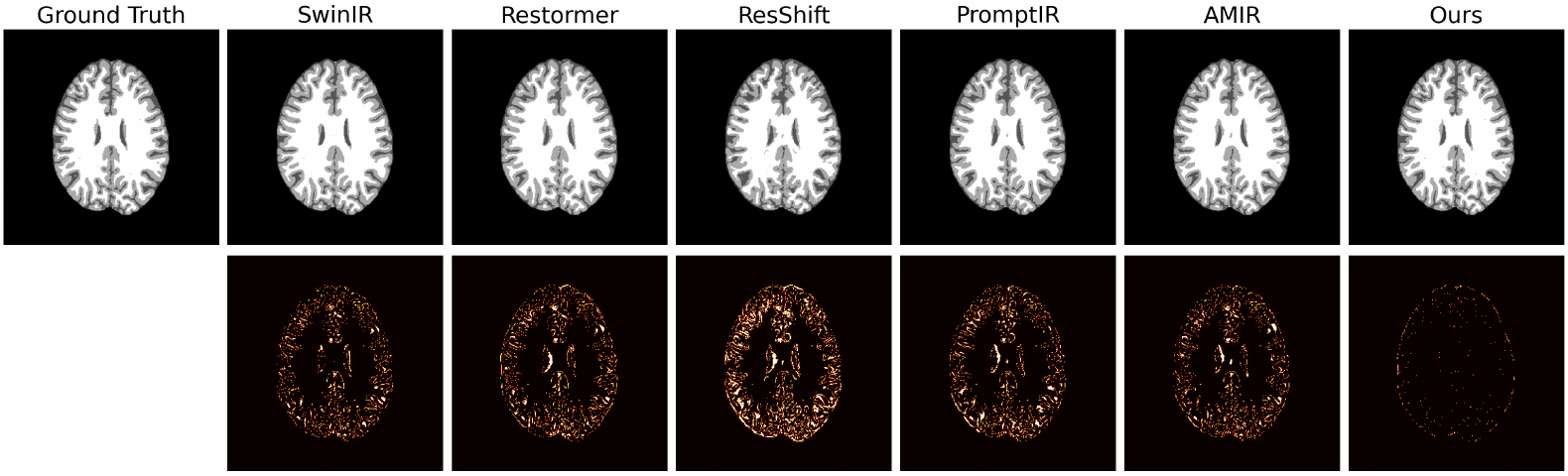}
    \caption{Visual comparison of automated brain tissue segmentations on representative MRI slices. The corresponding error maps highlight misclassified regions, indicating pixels where the predicted labels differ from the ground truth.}
    \label{fig:seg_visuals}
\end{figure}

As summarized in Table~\ref{tab:segmentation_results_pm_side}, our method consistently yielded the highest segmentation accuracy across all tissue classes. 
Relative to the next best-performing technique, our approach improved the mean Dice scores by approximately $1.70\%$ for CSF, $1.53\%$ for GM, and $0.94\%$ for WM. 
These quantitative gains are visually supported by the examples in Figure~\ref{fig:seg_visuals}, where our corrected images produce cleaner tissue boundaries and fewer misclassifications. 
Together, these results demonstrate that the proposed motion correction framework not only enhances image fidelity but also leads to more reliable downstream quantitative analyses in neuroimaging.

\begin{table}[htbp]
    \centering
    \caption{Quantitative evaluation of brain tissue segmentation performance on images corrected by different motion correction methods. CSF: Cerebrospinal Fluid. GM: Grey Matter. WM: White Matter.}
    \label{tab:segmentation_results_pm_side}
    \begin{tabular}{@{}lccc@{}}
        \toprule
        \multirow{2}{*}{Method} & \multicolumn{3}{c}{Mean Dice $\pm$ Std \%}  \\
        \cmidrule(lr){2-4} 
        & CSF & GM  & WM  \\
        \midrule
        SwinIR    & $88.94 \pm 1.24$ & $90.41 \pm 1.07$ & $94.44 \pm 0.54$ \\
        Restormer & $89.37 \pm 1.33$ & $91.38 \pm 0.74$ & $95.14 \pm 0.38$ \\
        ResShift  & $79.42 \pm 2.15$ & $82.67 \pm 1.35$ & $90.76 \pm 0.35$ \\
        PromptIR  & $90.07 \pm 0.94$ & $91.67 \pm 0.77$ & $95.30 \pm 0.41$ \\
        AMIR      & $90.47 \pm 0.93$ & $91.91 \pm 0.80$ & $95.42 \pm 0.37$ \\
        Ours      & $92.17 \pm 2.33$ & $93.44 \pm 2.08$ & $96.36 \pm 1.10$ \\
        \bottomrule
    \end{tabular}
\end{table}

\section{Conclusion}
\label{sec:conclusion}

We presented a unified framework for robust motion artifact correction in multi-contrast MRI, integrating text-informed contrast disentanglement with severity-aware adaptive experts. This design effectively addresses the dual challenges of contrast variability and motion severity, yielding significant improvements in image quality and downstream segmentation accuracy. Importantly, the framework demonstrated strong robustness on real-world data with severe artifacts, underscoring its translational potential.

Nevertheless, several limitations remain. First, the reliance on simulated motion for training may not fully capture the diversity of clinical artifacts, particularly affecting the generalization of the severity assessment module. Second, the current 2D-based composite quality score may lack sensitivity to localized degradations within 3D volumes, potentially limiting fine-grained severity estimation. These factors highlight the need for further methodological refinement.

Future work will focus on three directions: (i) large-scale, multi-center clinical validation with naturally occurring motion and domain adaptation strategies; (ii) extension to full 3D volumetric correction paired with 3D-aware quality metrics; and (iii) adaptation of the framework to other artifact types and degradations in medical imaging. 

In summary, by combining domain-specific acquisition knowledge with adaptive deep learning strategies, our framework advances the state of the art in MRI motion correction and provides a promising foundation for clinically reliable diagnostic imaging.

\appendix
\bibliographystyle{plain}
\bibliography{refs}




\end{document}